\documentclass[twocolumn,showpacs,prc,amsmath,amssymb,superscriptaddress]{revtex4}
\usepackage{graphicx}% Include figure files
\usepackage{dcolumn}% Align table columns on decimal point
\usepackage{bm}% bold math
\usepackage{epsfig}

\begin{document}

\preprint{APS/123-QED}

\title{Transient and quasi-stationary dissipative effects in the fission flux 
across the barrier in 1 A GeV $^{238}$U on deuterium reactions}
% Force line breaks with \\

\author{J.~Benlliure}
\email{j.benlliure@usc.es}
\affiliation{Universidad de Santiago de Compostela, E-15782 Santiago de 
Compostela, Spain}%Lines break automatically or can be forced with \\
\author{E.~Casarejos}%
\affiliation{Universidad de Santiago de Compostela, E-15782 Santiago de 
Compostela, Spain} 
\author{J.~Pereira}
\altaffiliation[Present address: ]{NSCL, Michigan State University, East Lansing, Michigan 48824, USA}
%\altaffiliation[Present address: ]{National Superconducting Cyclotron Laboratory, Michigan State University, East Lansing, Michigan 48824, USA}
\affiliation{Universidad de Santiago de Compostela, E-15782 Santiago de 
Compostela, Spain}
\author{K.-H.~Schmidt}
\affiliation{Gesellschaft f\"{u}r Schwerionenforschung,  D-64291
Darmstadt, Germany}

\date{\today}% It is always \today, today,
             %  but any date may be explicitly specified

\begin{abstract}
Isotopic cross sections of all projectile residues with Z above 23 
produced in collisions induced by $^{238}$U at 1 A GeV on deuterium have been 
measured. The isotopic distributions reflect the role of evaporation and 
fission in the formation process of these nuclei. The comparison of the 
measured cross sections with Monte-Carlo de-excitation codes including an 
analytical description of the dynamics of fission shows the sensitivity of the 
data to nuclear dissipation. Moreover, the large excitation-energy range 
covered in this experiment together with the high accuracy of the measured 
cross sections allowed to clearly separate and quantify the role of transient 
and quasi-stationary dissipative effects in the fission-decay width.
\end{abstract}

% PACS, the Physics and Astronomy
                             % Classification Scheme.
%\keywords{Suggested keywords}%Use showkeys class option if keyword
                              %display desired
\maketitle

\section{Introduction}
Intense effort is being invested to improve our understanding of nuclear dynamics, e.g. the evolution of colliding nuclear systems and the fission process. Most successfully used theoretical approaches consider these processes in terms of a few macroscopic collective variables, representing the shape of the system, and the individual motion of non-interacting nucleons, describing the intrinsic excitation. The dynamics of the system is described by transport equations of the Fokker-Planck or the Langevin type in a non-equilibrium statistical approach. A fundamental nuclear property, appearing as a parameter in these models, is the viscosity, respectively the dissipation strength, which measures the magnitude of the coupling between intrinsic and collective motion. This key property in particular governs the relaxation phenomena and diffusion processes of the nuclear system. Dissipation plays an essential role in the calculation of the formation cross sections of super-heavy nuclei (e.g. \cite{Abe02}) and in a consistent description of the fission process (e.g. \cite{Nad02}). Still, the magnitude of the dissipation strength and in particular an eventual dependence on temperature and/or deformation is under debate. Different theoretical models (one-body dissipation: \cite{Ran80}, linear response: \cite{Hof97}) yield diverging results. Moreover, quantum-mechanical effects, which go beyond these models, enter more and more into the discussion (e.g. \cite{Aba01,Cap02,Tak04,Rad05}). In this situation, it is of prime importance to improve our empirical knowledge in this field by well adapted experimental approaches.

The fission process has probably been the most successful tool for investigating nuclear dissipation. It is characterized by a clearly observable large-scale collective motion, ending up in a binary decay of the system, which is in competition with statistical decay by sequential evaporation of nucleons and light clusters. Theoretically, one expects several phenomena to appear in fission due to nuclear viscosity: Firstly, the system might show relaxation effects while establishing the quasi-equilibrium distribution in the shape coordinates. This phenomenon was first studied by Grang\'e, Jung-Qing and Weidenm\"uller \cite{Gra83} on the basis of the Fokker-Planck equation. As a consequence, the onset of the flow over the fission barrier is delayed by the transient time. The system may then cool down by evaporation, reducing the fission probability. Transient effects are expected to manifest only under specific experimental conditions \cite{Jur05a}. Secondly, when quasi-equilibrium is established, the quasi-stationary flow over the fission barrier is reduced with respect to the prediction of the Bohr-Wheeler transition-state model \cite{Boh39}. This reduction has already been deduced by Kramers \cite{Kra40}, who derived an analytical solution of the Fokker-Planck equation for an inverted parabolic potential. These two phenomena, which influence the flow over the fission barrier, will be studied in the present work. Thirdly, another consequence of dissipation is an increase of the dynamical saddle-to-scission time in the quasi-stationary flow. 

From the experimental side, there exist a number of observations that support the dynamical nature of the fission process. The anomalously enhanced pre-scission neutron multiplicities in fission induced in heavy-ion collisions \cite{Gav81,Hil92} have been interpreted as a signature of the delay of fission at high excitation energies. Other evidences for dissipative effects were obtained from the analysis of gamma-rays emitted during the de-excitation of the GDR \cite{Pau94} or directly measuring the fission time using crystal blocking techniques \cite{Gib75,Gon02}. From these experiments, the increase of the dynamical saddle-to-scission time due to dissipative effects appears to be a well-established phenomenon. 

The experimental proof for dissipative effects on the flow over the fission barrier remains an important subject of intensive research. In contrast to heavy-ion fusion-fission reactions, which suffer from complex initial conditions, e.g. broad angular-momentum distributions and large shape distortions, nucleon-induced reactions or very peripheral nucleus-nucleus collisions at relativistic energies seem to be best suited for these studies.

Recently, a novel experimental approach has been introduced \cite{Ben02,Jur04b}, which is based on the detection of fission products from nucleon-induced reactions or very peripheral nucleus-nucleus collisions at relativistic energies in inverse kinematics. The advantage of these reaction mechanisms is that the excited fissioning nucleus is produced with well defined initial conditions. In particular, these reactions lead to almost undeformed nuclei. In addition, they cover a large range in excitation energy while introducing only moderate angular momenta \cite{Jon97}. In these experiments, reaction residues with nearly projectile velocity are detected in forward direction. The inverse-kinematics conditions allow for identifying the fission products in atomic and mass number. These experiments were designed to be particularly sensitive to the time needed by the highly excited system to reach quasi-equilibrium. In these works, the fission cross sections, the charge distribution \cite{Jur04b}, and the isotopic distribution \cite{Ben02} of fission residues have been used as signatures of the fission dynamics. Evidences for relaxation effects leading to transient times in the order of a few 10$^{-21}$ s were deduced \cite{Ben02,Jur04b}, which corresponds to conditions close to the minimun possible transient time realized at critical damping \cite{Jur05b}.

Since several years, also experiments on light charged-particle-induced reactions, e.g. \cite{Sch97,Tis05} in conventional kinematics have been performed. In this case, evaporative neutrons and light charged particles were measured, from which initial thermal excitation energies were evaluated event by event. In addition, the fission fragments were registered, however, with a rather low resolution in mass. In a very recent letter \cite{Tis05}, the measured fission probabilities of highly excited systems in 2.5-GeV proton-induced reactions were compared with model calculations. From this comparison the authors could not find any indication for dynamical effects on the flux over the fission barrier, since the data were fully explained by the statistical Bohr-Wheeler approach.

Thus, the findings deduced from the signatures of these two different experiments are in clear contradiction. In addition, the conclusions of Ref. \cite{Tis05} are rather intriguing by themselves. In particular, the following two statements of the paper are hardly compatible in view or our present theoretical understanding of the fission process: On the one hand, the protons are expected to induce only little collective excitations by angular momentum, shape distributions and compression. On the other hand, it is claimed that fission is in competition with particle evaporation immediately after the nuclear-collision stage of the reaction. It seems that the deduced immediate onset of fission requires a very fast population of the transition states above the fission barrier in a time well below the minimum transient time imaginable. Thus, the conclusions drawn in Ref. \cite{Tis05} seem to require a revised understanding of nuclear dynamics. If we admit that the inertia of the nuclear system is rather well established, it seems that the validity of the non-equilibrium statistical approach in terms of the transport equations of the Fokker-Planck or the Langevin type must be questioned.

The present work is intended to contribute to the solution of these problems. We investigate the reaction $^{238}$U on deuterium at 1 A GeV in inverse kinematics, taking advantage of the unique experimental installations of GSI. However, this time we use a different observable, the isotopic production cross sections of evaporation residues. Due to the high fissility of the projectile, the yields of evaporation residues are expected to strongly depend on the fission process and in particular on transient and quasi-stationary dissipative effects on the fission flux across the barrier. In addition, the large excitation-energy range covered in the collisions induced by $^{238}$U on deuterium can be sorted according to the mass loss of the final residues with respect to the projectile. Because of the narrow mass range produced in the collision stage of the reaction, the mass loss is dominated by the evaporation process, which is directly related to the excitation energy induced in the collision. One can then investigate the influence of transient and quasi-stationary dissipative effects in the fission flux as a function of the excitation energy involved in the fission process. Thus, the information on the physics of the fission process provided by the present experiment is very similar to the one obtained in Ref. \cite{Tis05}, although it is based on different observables. In addition, the system is very similar, providing almost the same amount of centre-of-mass energy.

\section{Experiment and results}
\label{sec:1}

\begin{figure*}
  \begin{center}
    \leavevmode
    \includegraphics[width=14.0cm]{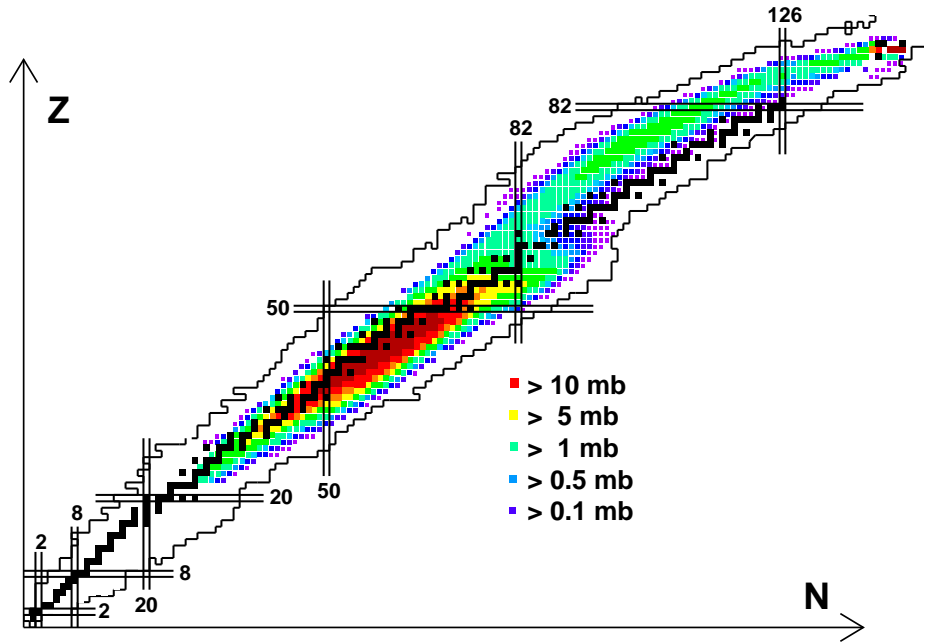}
    \caption{Two-dimensional cluster plot of all nuclei produced as projectile 
residues with a cross section larger than 100 $\mu$b in the reaction 
$^{238}$U+d at 1 A GeV presented on top of a chart of the nuclides. The colour 
scale represents the production cross section, the black squares correspond to 
stable isotopes, and the lines indicate the limit of the known nuclides.}
  \end{center}
\end{figure*}

The experiment was performed at GSI (Darmstadt) where the SIS synchrotron was used to produce a pulsed beam of $^{238}$U at 1 A GeV with a maximum intensity of 10$^7$ ions s$^{-1}$, a pulse length of 3 s and a repetition cycle of 7 s. This beam impinged onto a liquid-deuterium target with a thickness of 200 mg/cm$^2$. The cylindrical target container was 1 cm long and 3 cm in diameter with titanium windows of 18 mg/cm$^2$ in total.

Due to the inverse kinematics, the projectile residues produced in the reaction were emitted in forward direction and could be magnetically analysed in-flight with the fragment separator FRS \cite{Gei92}. This is an achromatic two-stage zero-degree spectrometer with a dispersive intermediate image plane. Every stage is equipped with two dipoles and a set of quadrupoles in front and behind each dipole. The resolving power of this device is B$\rho$/$\Delta$B$\rho$ $\approx$ 1500 with an acceptance of 15 mrad around the beam axis in angle and $\pm$1.5\% in momentum.

In order to isotopically identify all the transmitted nuclei, their horizontal positions at the intermediate and final image planes, their velocity and their energy loss in a given gas volume were measured with different detectors. Two plastic scintillators, about 20 cm long, 8 cm height and 5 mm thick, equipped with two photomultipliers at both extremes, were placed at the intermediate and final image planes to provide the position and time-of-flight measurements \cite{Vos95}. The atomic number of the nuclei was obtained from their energy-loss measurement in two multi-sampling ionisation chambers \cite{Pfu94} located behind the spectrometer. Finally, a set of multi-wire chambers placed at the different image planes of the spectrometer was used for calibration at the beginning of the experiment. In addition, and in order to identify the atomic number of heavy reaction residues with Z above 70, a velocity degrader \cite{Sch87} at the intermediate image plane of the spectrometer was used. The final resolution (FWHM) achieved for the mass and charge separation was A/$\Delta$A $\approx$ 400 and $\Delta$Z $\approx$ 0.4, respectively. The mass and charge identifications of the reaction residues were obtained by using the primary beam as reference and the specific pattern of light nuclei when represented on a two-dimensional spectrum A/Z versus Z \cite{Ric04}.

Due to the limited acceptance of the spectrometer, more than 50 different magnetic tunings of the dipoles of the FRS were needed to completely measure the momentum distributions of all nuclear species produced in the reaction $^{238}$U on deuterium at 1 A GeV with a production cross section above 100 $\mu$b and Z$>$23. Using the information provided by the beam-monitor detector SEETRAM \cite{Jun96}, the yields measured in different magnetic tunings were normalised and combined to produce the corresponding momentum distribution for each nuclide. The integral of these momentum distributions normalised to the
number of atoms in the target and corrected by the detection efficiency, angular transmission and secondary reactions provided the production cross sections of all measured nuclei. A detailed description of the experiment and the data analysis for a similar experiment can be found in references \cite{Tai03,Ber03}.

The results of the experiment are summarised in Fig.1. In this figure, we represent in a two-dimensional cluster plot all the projectile residues produced in the reaction $^{238}$U on deuterium at 1 A GeV, with a cross section larger than 100 $\mu$b and Z above 23, on top of a chart of the nuclides. More than 1400 different nuclides have been identified, and their cross sections have been determined with an accuracy between 10\% and 20\%. In the figure we can clearly observe two different groups of nuclides. The upper part of the chart of the nuclides is populated by mostly neutron-deficient nuclides covering a large range in charge from the projectile nucleus down to the charge 65. On average, for a given element, the isotopic distribution covers around 20 different nuclides with a maximum production cross section around a few millibarns. These nuclei are produced in a nucleon and/or cluster evaporation process from the excited projectile pre-fragments produced in the interaction between the projectile and the target nuclei. This process leads to residual nuclei lighter than the projectile with an isotopic composition determined by the competition between neutron and proton evaporation and the initial excitation energy induced in the collision. Since the evaporation of protons and neutrons is governed not only by the respective binding energies but also by the proton Coulomb barrier, the equilibrium between both processes is reached at the left of the stability line, along the so called ``evaporation corridor'' \cite{Duf82}, defining the ridge line of the nuclide distributions of evaporation residues. The initial excitation energy determines the length of the evaporation chain and consequently the mass of the final residue. Since the number of nucleons removed from the projectile in direct collisions with the target deuteron is expected to be rather small, the mass of the evaporation residues is an observable which is well correlated with the excitation energy induced in the reaction. This important aspect will be discussed later in further detail.

The second group of nuclei observed in Fig. 1 corresponds to medium-mass fragments with atomic numbers between 23 and 65. In this case, the isotopic distributions are broader, populating on average 25 isotopes. The distributions are centred to the right of the stability line, and the production cross sections are much larger than the ones observed for the evaporation residues. These nuclei are expected to be produced by fission of the projectile pre-fragments emerging from the interaction between the $^{238}$U projectiles with deuterium. The neutron excess of the projectile pre-fragments is preserved in the fission process. However, when the system crosses the fission saddle at high excitation energies, this neutron excess is partially lost by post-saddle neutron evaporation, both from the system on the way to scission and from the fragments, shifting and broadening the final isotopic distribution of fission residues. The observed mass distribution of fission residues is mostly single-humped, in contrast to the well known double-humped distribution of residues produced in the low-energy fission of $^{238}$U, indicating in our case that fission takes place at high excitation energy, where shell effects are washed out. Nevertheless, the most neutron-rich fission residues show a contribution by a double-humped component resulting from low-energy fission induced in very peripheral reactions.

\begin{figure}
  \begin{center}
    \leavevmode
    \includegraphics[width=9cm]{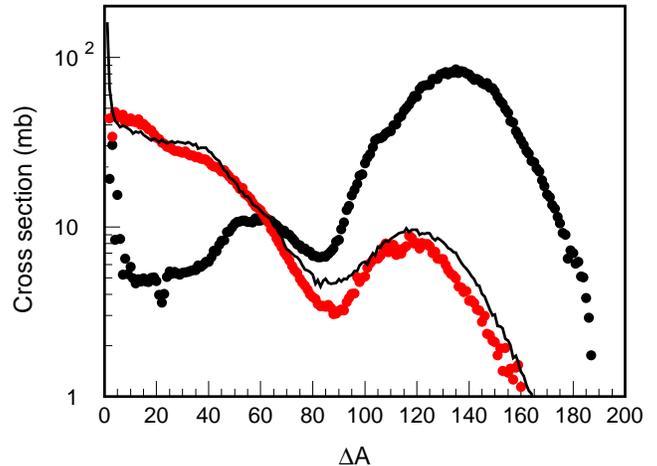}
        \caption{Isobaric distribution of projectile residues produced in 
reactions induced by $^{238}$U(1 A GeV) (black dots) and $^{208}$Pb(1 A GeV) 
(grey dots) \cite{Enq02} on deuterium as a function of the mass loss of the 
final residues with respect to the mass of the projectile. The isobaric 
distribution of residues produced in the reaction $^{208}$Pb(1 A GeV) on 
deuterium is compared to a model calculation (solid line) performed with the 
Isabel intra-nuclear cascade code \cite{Yar81} coupled to the ABLA evaporation 
code \cite{Jun98,Ben98} including a dynamical description of fission according 
to reference \cite{Jur05a}.}
  \end{center}
\end{figure}

From the analysis of Fig. 1 we can conclude that fission is the dominant de-excitation channel in the reaction $^{238}$U on deuterium at 1 A GeV. A similar conclusion can be drawn from Fig. 2. In this figure, we represent the isobaric distribution of projectile residues measured in the reaction $^{238}$U(1 A GeV) on deuterium (black dots) compared to the one obtained for the reaction $^{208}$Pb(1 A GeV) on deuterium (grey dots) \cite{Enq02}. We can also identify the two groups of residues produced in evaporation and fission processes, respectively. Although the excitation energy induced in both reactions should be similar, fission seems to be the dominant reaction channel in the collisions induced by $^{238}$U, while in the case of the reactions induced by $^{208}$Pb we arrive to the opposite conclusion. This result can be attributed to the different fissilities of the pre-fragments produced in both reactions. Since the fission component for the $^{208}$Pb system represents less than 15\% of the total cross section, and the total cross sections for the reactions $^{238}$U(1 A GeV) and $^{208}$Pd(1 A GeV) on deuterium differs by less than 15\%, we can conclude that the differences observed in Fig. 2 for the isobaric distributions of evaporation residues in both reactions are mostly due to the fission process in the reactions induced by $^{238}$U at 1 A GeV on deuterium.

\section{Analysis of dissipative effects}
\label{sec:2}

\subsection{Model description}

A deeper and more quantitative interpretation of our data requires the use of model calculations. In order to describe the interaction of deuterium with $^{238}$U at relativistic energies, we used an intra-nuclear cascade code coupled to a de-excitation code considering both nucleon and/or cluster evaporation and fission. After benchmarking different intra-nuclear-cascade models, the most consistent description of the available data on $^{238}$U + $^2$H and $^{208}$Pb + $^2$H \cite{Enq02}, which are characterised by substantially different fissilities, was obtained with the Isabel code of Yariv and Fraenkel \cite{Yar81}. The evaporation and fission processes were described using the Monte-Carlo code ABLA. In this code the evaporation of nucleons and alpha clusters is described according to the Weisskopf formalism including a consistent description of level densities where shell, pairing and collective effects are considered following reference \cite{Jun98}. The fission width is calculated using the statistical approach of Bohr and Wheeler $\Gamma_{BW}$~\cite{Boh39}, corrected for quasi-stationary and transient dissipative effects, which describe the dynamical nature of fission according to:

\begin{equation}
\Gamma_{f}(t) = \Gamma_{BW} \cdot K \cdot f(t)
\label{eq_1}
\end{equation}

In this equation $K$ represents to the so-called Kramers coefficient which provides the correction one has to apply to the statistical fission width in order to obtain the quasi-stationary solution of the Fokker-Plack equation describing the fission flux across the barrier \cite{Kra40}. The function $f(t)$ describes the time dependence of the fission width resulting from the time-dependent solution of the Fokker-Planck equation proposed by Grang\'e, Jun-Qing and Weidenm\"uller \cite{Gra83}. The analytical approach of Ref. \cite{Jur05a} to Eq. \ref{eq_1} makes it possible to avoid the computational effort required to solve numerically the Fokker-Planck equation on a event-by-event basis. 

In Eq. \ref{eq_1} the Kramers factor is defined as:

\begin{equation}
K = \left\{ \left[ 1 + \left(\frac{\beta}{2\omega_o} \right)^2 \right]^{1/2} - \frac{\beta}{2\omega_o} \right\}
\label{eq_2}
\end{equation}

where $\beta$ is the reduced dissipation coefficient, and $\omega_o$ corresponds to the frequency of the harmonic oscillator describing the inverted potential at the fission barrier. In our Monte-Carlo code we use different analytical approximations to describe the time dependence of the fission width $f(t)$ according to Ref. \cite{Jur05b}.

The main consequence of Eq. \ref{eq_2} is that the quasi-stationary and transient effects that appear in the dynamical description of the fission flux across the barrier lead to a reduction of the fission width compared to the value obtained with the statistical model.

Besides nuclear dissipation, the fission barriers and even more the level densities are the ingredients of the model description, which have the most important influence on the fission probabilities at high excitation energies and thus should be considered with special care. In addition to the arguments given in \cite{Ben02}, which already justified the choice of Ref. \cite{Sie86} and \cite{Ign75} for the description of fission barriers and level densities, the deformation dependence of the level-density parameter, which is the basis for their ratio at the transition state and ground deformation a$_f$/a$_n$ used in our calculation, has been investigated recently on the basis of the Yukawa-plus-exponential description of the nuclear properties \cite{Kra99,Kar03}. Also this theoretical study has essentially confirmed the validity of Ref. \cite{Ign75}, which we use in our calculation.

Finally, the isotopic distribution of fission residues is described following the model of reference \cite{Ben98}. In addition, the ABLA code includes a breakup channel which sets in when the nuclear temperatures reaches 5 MeV \cite{Sch02}.

An example of the results obtained with this model is shown in Fig. 2 (solid line) for the reaction $^{208}$Pb(1 A GeV) on deuterium. This reaction can be considered as an optimum case to benchmark the intra-nuclear and the evaporation code, since in this case the role of fission is minor. Moreover, the description of the level densities is simpler since only few nuclei will feel the Z=82 shell, and the collective enhancement for the deformed nuclei is expected to be almost the same for the different de-excitation channels, e.g. in the ground state for particle emission and at the saddle for fission. In this calculation we described the fission dynamics by means of the analytical approximated solution of the time-dependent Fokker-Planck equation, proposed by B. Jurado and collaborators~\cite{Jur05a,Jur05b}, with a reduced dissipation coefficient $\beta=2 \cdot 10^{21}~s^{-1}$. The good description of the production cross sections provided by the code for both, evaporation and fission residues, validates the intra-nuclear cascade model as well as the description of the evaporation channels and the breakup, which plays a role only for the lightest evaporation residues.

\subsection{Discussion}

In order to better understand the general trends of the reaction $^{238}$U(1A GeV)+d, we show in Fig.~3 the calculated excitation energy of the initial pre-fragments as a function of the mass-loss of the final residue ($\Delta$A) with respect to the projectile. This calculation was done with the same description of the fission width used for the calculation of Fig.~2.

%%%NEW FIGURE%%%%Take care with the new numbering!!!!
 \begin{figure}
  \begin{center}
    \leavevmode
    \includegraphics[width=8cm]{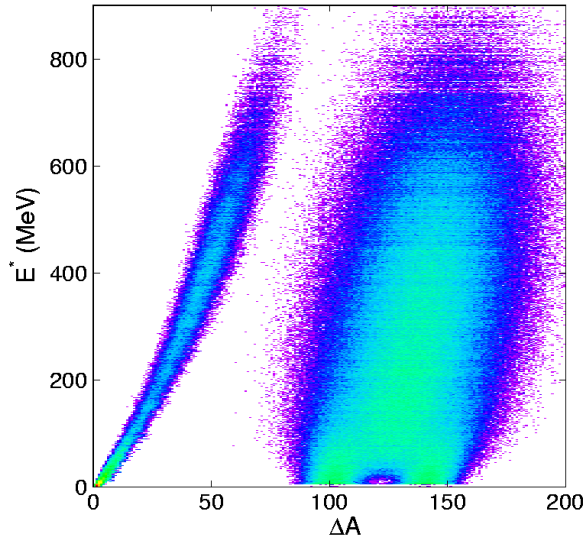}
        \caption{Excitation energy of the initial pre-fragments as a function
of the mass loss of the final residue for the reaction $^{238}$U (1A GeV)+d.}
\end{center}
\end{figure}

As can be seen, only evaporation residues ($\Delta$A \(<\) 90) exhibit a clear correlation between their final masses and the initial excitation energies of the decaying pre-fragments. Applying different intra-nuclear cascade codes \cite{Yar81,Bou02}, we found that this correlation is a common feature, which quantitatively varies only little for the different codes. We understand this result as a consequence of a rather strict correlation between the initial excitation energy and the mass loss in the de-excitation stage, the mass loss in the intra-nuclear cascade stage being comparably small. In the light of this figure, we can re-examine the residue productions shown in Fig.~2 for the reaction $^{238}$U (1A GeV)+d. 

Considering the existing correlation between the measured mass loss of the final evaporation residues with respect to the projectile and the initial excitation energy of the pre-fragments and the differences in the yields of evaporation residues in collisions induced by $^{238}$U and $^{208}$Pb, we arrive to an interesting observation: The fission channel seems to dominate the de-excitation process of pre-fragments produced in the reaction $^{238}$U(1 A GeV) on deuterium at low and moderate excitation energies. (Note that Fig. 2 provides an energy scale to the ordinate of Fig. 3). However, the similar production cross sections of light evaporation residues ($\Delta$A $\approx$ 70) observed in both reactions clearly indicate that the fission of $^{238}$U pre-fragments is suppressed at high excitation energies. This is a surprising result that clearly contradicts the expectations due to the statistical description of the fission process according to the model of Bohr and Wheeler \cite{Boh39}. This inhibition of fission at high excitation energies would qualitatively fit with the dynamical interpretation of fission as proposed by Kramers \cite{Kra40} and later on by Grang\'e, Jun-Qing and Weidenm\"uller \cite{Gra83}.

 \begin{figure}
  \begin{center}
    \leavevmode
    \includegraphics[height=7cm]{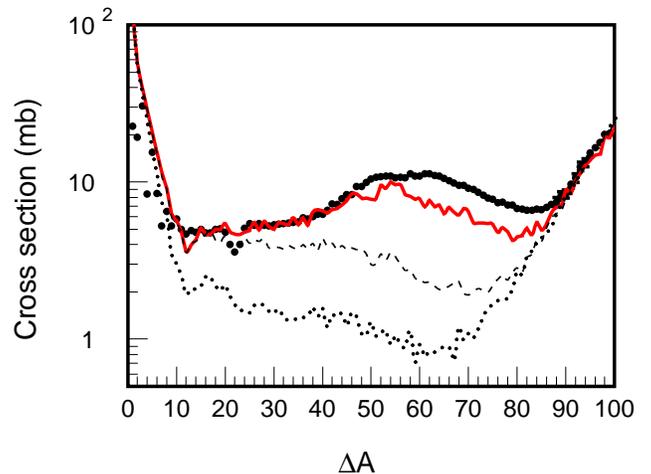}
        \caption{Measured isobaric distribution of evaporation residues
produced in the reaction $^{238}$U(1 A GeV)+d (dots), compared with different
calculations performed with the Isabel intra-nuclear cascade code coupled to
the ABLA evaporation code (see text for details).}
  \end{center}
\end{figure}

Once we have validated most of the ingredients in our code we can concentrate on the fission channel which is the dominant one in the reaction induced by $^{238}$U. In Fig. 4, we compare the measured isobaric distribution of evaporation residues produced in the reaction $^{238}$U(1 A GeV)+d (dots) with different model calculations. The dotted line was obtained using the purely statistical model of Bohr and Wheeler \cite{Boh39}. As can be seen, this model clearly overestimates the fission channel depopulating the production of evaporation residues. Better agreement with the measured data is obtained when the dynamics of fission is considered in the calculation. The dashed line corresponds to a calculation where the fission width is evaluated using the Kramers factor described by equation \ref{eq_2} with a reduced viscosity parameter of $\beta$ = 2$\cdot$10$^{21}$ s$^{-1}$. However, a calculation considering not only the stationary effects in the fission flux by means of the Kramers factor with  $\beta$ = 2$\cdot$10$^{21}$ s$^{-1}$, but also the transient effects due to the time dependence of the fission width (solid line) provides the best description of the data. In this case, the time dependence of the fission width has been calculated according to the approximated solution of the Fokker-Planck equation proposed by Jurado and collaborators \cite{Jur05a,Jur05b}.

The fair agreement observed in Fig.~4 between our calculations (solid line) and the experimental data (dots) is consistent with previous investigations where different observables and reactions have been used. In particular, partial fission cross sections and final charge distributions of fission residues produced in the reaction $^{238}$U(1 A GeV)+CH$_2$  \cite{Jur04b}, total fission cross section and the isotopic distribution of fission residues in collisions induced by $^{197}$Au on hydrogen at 800 A MeV \cite{Ben02} and the present isotopic distributions of evaporation residues produced in the reaction $^{238}$U at 1 A GeV on deuterium, are coherently reproduced by these calculations. 

However, until now, none of the observables used as signatures of the fission dynamics allowed us to characterise and quantify independently the role of transient and quasi-stationary dissipative effects in the fission flux at small deformation. In principle one would expect low-energy fission to be sensitive to quasi-stationary effects, characterised by the reduced dissipation coefficient in Kramers factor, and not to transient effects, which are expected to be much shorter than the statistical fission time at low excitation energy. On the other hand, as the excitation energy increases, the statistical fission time approaches the transient time, and the fission process becomes sensitive to this latter effect. The strong correlation between the final mass of the evaporation residues and the initial excitation energy, together with the large excitation-energy range covered with the reaction $^{238}$U at 1 A GeV on deuterium allowed us to analyse our data following these ideas.

 \begin{figure*}
  \begin{center}
    \leavevmode
%Figure 5%
    \includegraphics[height=8cm]{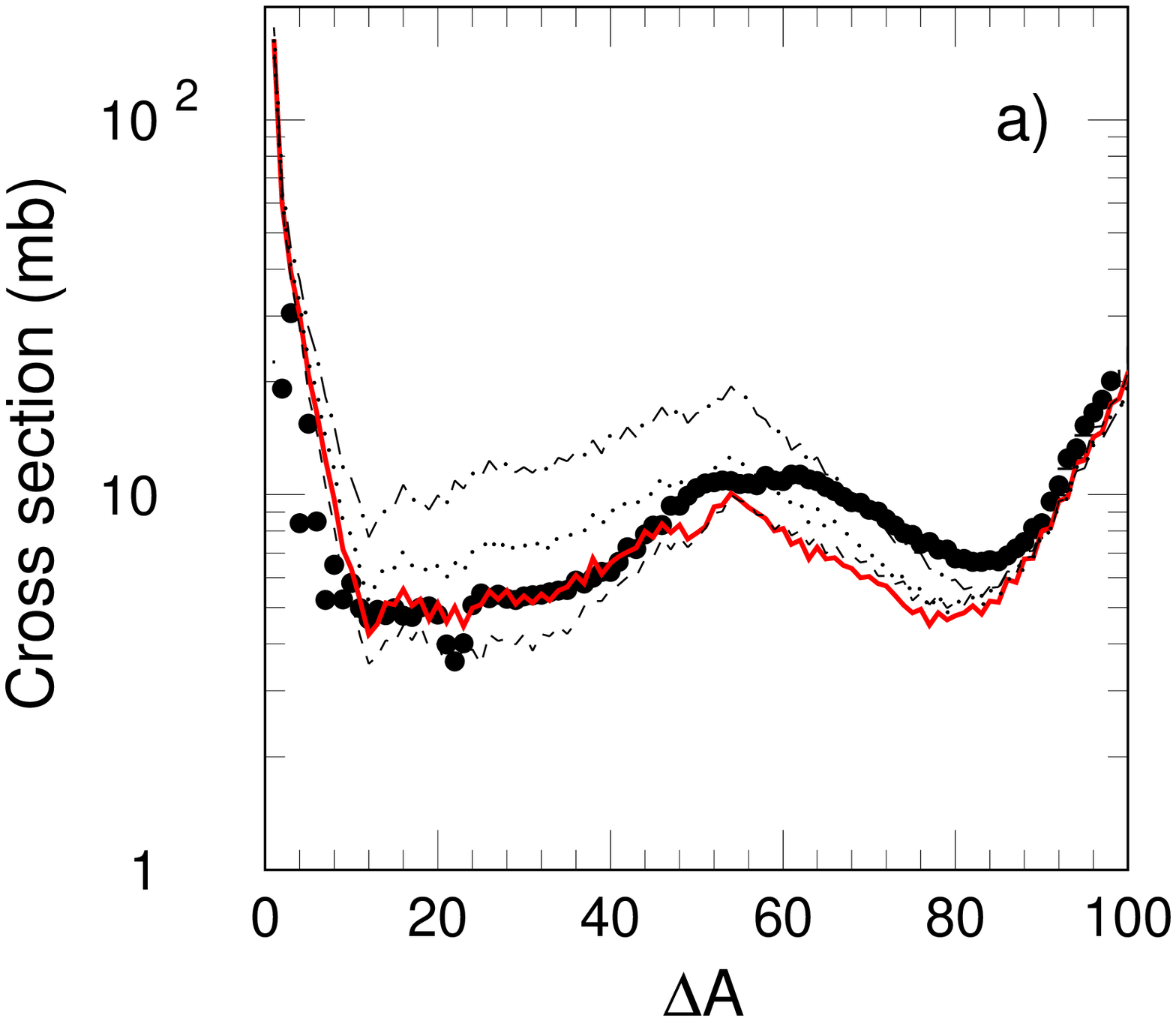} \includegraphics[height=8cm]{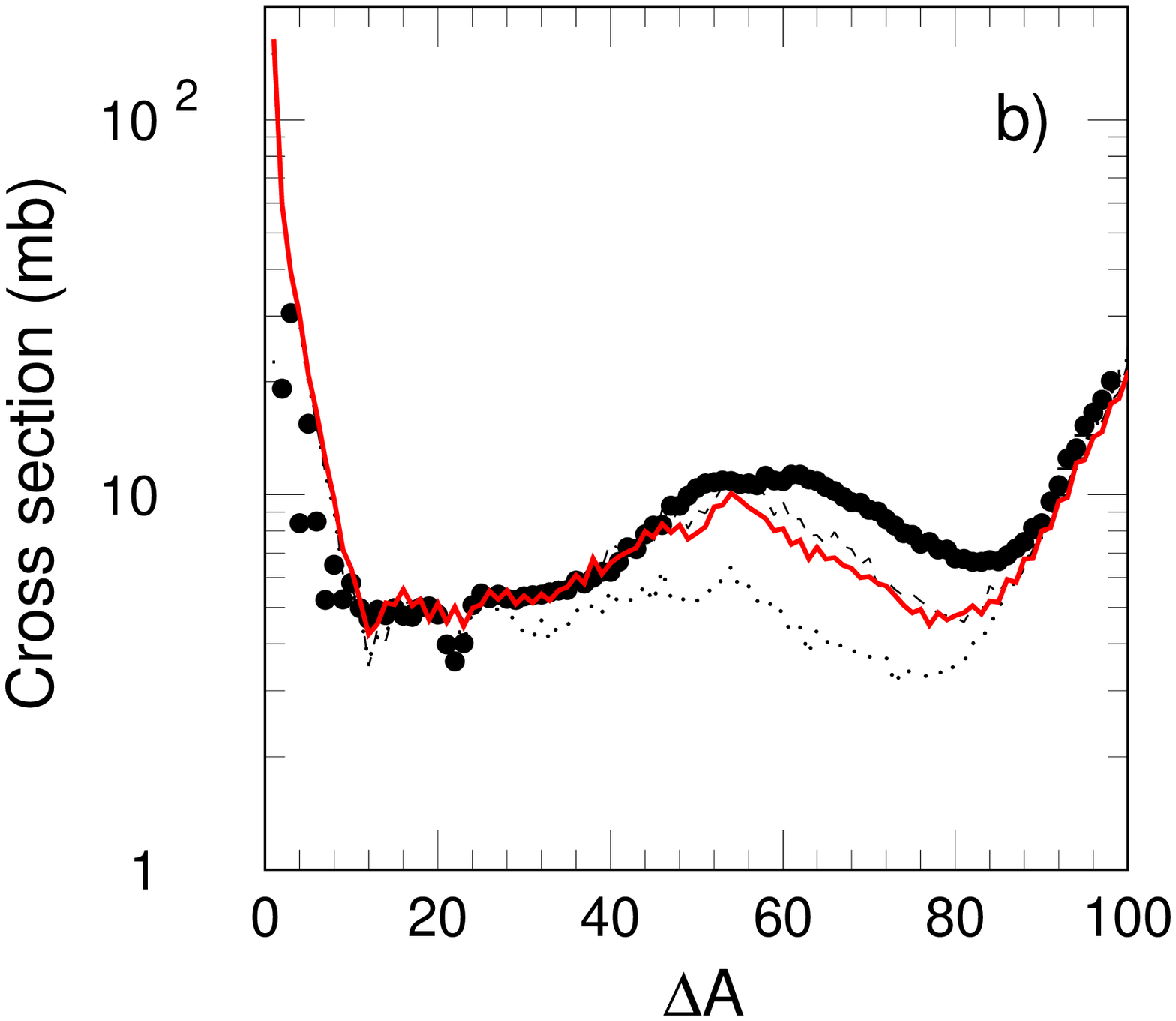}
        \caption{Isobaric distribution of evaporation residues produced in the
reaction $^{238}$U+d at 1 A GeV (dots). In the left panel, the data are
compared to calculations using different values of $\beta$:
1$\cdot$10$^{-21}$s$^{-1}$ (dashed line), 2$\cdot$10$^{-21}$s$^{-1}$
(solid line), 3$\cdot$10$^{-21}$s$^{-1}$ (dotted line) and
5$\cdot$10$^{-21}$s$^{-1}$ (dashed-dotted line), in all these calculations the
time dependence of the fission width follows the analytical solution of the
FPE. In the right panel, we compare calculations with
$\beta$ = 2$\cdot$10$^{-21}$s$^{-1}$, but different time dependences of the
fission width: an exponential in-growth function (dotted line), a step function
(dashed line) and the analytical solution of the FPE (solid line).}
  \end{center}
\end{figure*}

In Fig.~5 we show the isobaric distribution of evaporation residues produced in  the reaction $^{238}$U at 1 A GeV on deuterium compared with calculations using different reduced dissipation coefficients and descriptions of the time-dependent fission width. In the left panel of the figure we compare the measured data to calculations, where different values of the reduced dissipation coefficient $\beta$ have been used, $\beta$ = 1$\cdot$10$^{21}$ s$^{-1}$(dashed line), $\beta$ = 2$\cdot$10$^{21}$ s$^{-1}$(solid line), $\beta$ = 3$\cdot$10$^{21}$ s$^{-1}$ (dotted line) and $\beta$ = 5$\cdot$10$^{21}$ s$^{-1}$ (dotted-dashed line). In all these calculations the time dependence of the fission width has been described using the approximate solution of the Fokker-Planck equation proposed by Jurado et al. \cite{Jur05a,Jur05b}. These calculations clearly show how the measured yields of evaporation residues depend on the value of the reduced dissipation coefficient. Even more, this effect manifests mainly in heavy evaporation residues corresponding to the survival probability against fission at low or moderate excitations energies. As already stated, the value of the reduced dissipation coefficient $\beta$ = 2$\cdot$10$^{21}$ s$^{-1}$ reproduces the data fairly well.

In the right panel of Fig.~5, we compare the same data to another set of model calculations. This time we have fixed the value of the reduced dissipation coefficient to $\beta$ = 2$\cdot$10$^{21}$ s$^{-1}$ and we have used different models to describe the time dependence of the fission width. The result of the calculation using the approximate solution of the Fokker-Planck equation proposed by Jurado et al. \cite{Jur05a,Jur05b} is represented by the solid line. The dotted line corresponds to an exponential in-growth time dependence of the fission width according to the expression \cite{But91}

\begin{equation}
\Gamma_f(t) = K\cdot\Gamma_{BW}\cdot \left\{1-\exp(-t/\tau)\right\}
\label{eq_3}
\end{equation}

where $K$ is the Kramers factor, $\Gamma_{BW}$ the statistical fission width and $\tau$=$\tau_{trans}$/2.3 with $\tau_{trans}$ being the transient time defined as 90\% of the time the fissioning system needs to reach the stationary fission-decay width across the barrier. In this figure, the dashed line represents the result obtained with a time dependence of the fission width following a step function \cite{Ras91}.

\begin{equation}
\Gamma_f(t) = \left\{
\begin{array}{ll}
0 & ~~~~t < \tau_{trans}\\
K \cdot \Gamma_{BW} & ~~~~t \geq \tau_{trans}\\
\end{array}
\right.
\label{eq_4}
\end{equation}

From this comparison we can conclude that the time dependence of the fission width based on the approximated solution of the Fokker-Planck equation and the step function provide similar results, although the first one has a better physical justification. The exponential in-growth clearly overestimates the fission width, in particular at high excitation energies (short times) as was already pointed out in reference \cite{Jur05b}. In addition, we can also conclude that in contrast to the calculation performed with different values of  the reduced dissipation coefficient $\beta$, in this case the sensitivity of the data to the transient effects appears for large values of the mass loss of the residues corresponding to high excitation energies. Consequently, our data make it possible to decouple the role of transient and quasi-stationary dissipative effects in the fission flux. We describe the evaporation-residue cross sections for a large mass range with the same value of the reduced dissipation coefficient ($\beta$ = 2$\cdot$10$^{21}$ s$^{-1}$) fairly well. Slight deviations for the lightest residues might indicate an even stronger suppression of fission at the highest excitation energies, since the measured cross sections of these residues are slightly underestimated by the calculation. Before drawing more quantitative conclusions, improved theoretical calculations are planned to better describe the entrance-channel distribution of the pre-fragments in deformation space, which we assumed to be identical to the zero-point motion of the projectile around a spherical shape in the present analysis.

When we compare our conclusions with the ones drawn in Ref. \cite{Tis05}, we observe a severe discrepancy. While we find strong indications for dissipative effects in the magnitude of the evaporation-residue cross sections, the essentially complementary fission probabilities measured in \cite{Tis05} were fully reproduced by calculations with the purely statistical Bohr-Wheeler approach. For the system $^{238}$U + $^2$H, we noticed an essential difference in the ingredients of the model calculation, which probably explains the controversial conclusions: While we determine the appropriate value of a$_f$/a$_n$ according to the saddle-point deformation of each fissioning system following the description of Ignatyuk et al. \cite{Ign75} for the deformation dependence of the level density parameter, Tishchenko et al. \cite{Tis05} use a common value of a$_f$/a$_n$ = 1 for all fissioning systems produced in the collision of $^{238}$U with 2.5 GeV protons, which extend over a broad range of elements \cite{Jur04b}. With the option a$_f$/a$_n$ = 1, we also can reproduce the fission probabilities reported in Ref. \cite{Tis05} within the Bohr-Wheeler statistical approach fairly well without introducing any transient effects. Thus, the diverging findings of Tishschenko et al. and the present work for the fission of $^{238}$U by 2 GeV light charged particles seem to be traced back to different ingredients of the model calculations.

A discussion of the results for the lighter systems also studied in \cite{Tis05} is beyond the scope of the present work. We will address this interesting subject in the near future, when data from inverse-kinematics experiments for similar systems will be available. 

\section{Conclusion}

In this letter, we have shown that the accurate measurement of the isotopic production of reaction residues provides valuable information on the reaction mechanism. In particular, we have presented the measured isotopic cross sections of more than 1400 different projectile residues produced in the reaction $^{238}$U(1 A GeV) on deuterium with an accuracy between 10\% and 20\%. In this reaction, the dominant fission channel can be investigated from the measured fission residues or from the survival probability against fission, represented by the evaporation residues. However, the strong correlation existing between the mass loss of the evaporation residues and the initial excitation energy induced in the collisions offers the possibility to use these residual nuclei for investigating the excitation-energy dependence of the fission process. The comparison of these data with the evaporation residues produced in the reaction $^{208}$Pb(1 A GeV) on deuterium clearly shows the dominance of fission in the $^{238}$U-induced reaction. However, the fact that for large mass losses both reactions lead to the production of residues with similar cross sections represent a clear signature of the suppression of fission in the $^{238}$U-induced reactions at high excitation energies. Model calculations show that the measured data can only be understood when the dynamics of fission is considered, in agreement with previous works. However, the accuracy of the present data, together with the large excitation-energy range covered with this reaction made it possible to characterise and quantify independently the role of transient and quasi-stationary dissipative effects in the fission flux across the barrier. The cross sections in the whole mass range are fairly well reproduced with a single value of the dissipation coefficient $\beta$ = 2$\cdot$10$^{21}$ s$^{-1}$.

Discrepancies in the interpretation of similar data recently obtained by Tishchenko el al. \cite{Tis05} in direct kinematics were understood in terms of the different ingredients used in the model calculations. The use of appropiate values for the level density parameter and in particular its dependence with deformation seems to be crucial for the correct understanding of the fission dynamics at high excitation energy.

\begin{acknowledgments}
We thank B. Jurado, A. Kelic, F. Rejmund and C. Volant for valuable discussions as well as T. Enqvist for the careful reading of the manuscript. This work was partially supported by the European Commission through the HINDAS project under contract FIKW-CT-2000-031 and the Spanish MEC and XuGa under contracts FPA2002-04181-C04-01 and PGIDT01PXI20603PM, respectively.
\end{acknowledgments}

%\bibliography{apssamp}% Produces the bibliography via BibTeX.

\begin{thebibliography}{99}

\small

\bibitem{Abe02} Y. Abe, Eur. Phys. J. A 13 (2002) 143

\bibitem{Nad02} P. N. Nadtochy, G. D. Adeev and A. V. Karpov, Phys. Rev. C 65 (2002) 064615

\bibitem{Ran80} J. Randrup and W. J. Swiatecki, Annals of Physics 125 (1980) 193

\bibitem{Hof97} H. Hofmann, Phys. Rep. 284 (1997) 137

\bibitem{Aba01} G. Abal, Romanelli, A. C. Sicardi-Schifino, R. Siri and R. Donangelo, Nucl. Phys. A 683 (2001) 279 

\bibitem{Cap02} L. Capriotti, A. Cuccoli, A. Fubini, V. Tognetti and R. Vaia, Europhys. Lett. 58 (2002) 155 

\bibitem{Tak04} N. Takigawa, S. Ayik, K. Washiyama and S. Kimura, Phys. Rev. C 69 (2004) 054605  

\bibitem{Rad05} S. Radionov and S. Aberg, Phys. Rev. C 71 (2005) 064304

\bibitem{Gra83} P. Grang\'e, LiJun-Qing and H.A. Weidenm\"uller, Phys. Rev. C27 (1983) 2063

\bibitem{Jur05a} B. Jurado et al., Nucl. Phys. A 757 (2005) 329

\bibitem{Boh39} N. Bohr and J.A. Wheeler, Phys. Rev. 56 (1939) 426

\bibitem{Kra40} H.A. Kramers, Physika VII 4 (1940) 284

\bibitem{Gav81} A. Gavron et al. Phys. Rev. Lett. 47 (1981) 1255

\bibitem{Hil92} D.J.~Hilscher and H.~Rossner, Ann. Phys. Fr. 17 (1992) 471

\bibitem{Pau94} P. Paul and M. Th\"onnessen, Ann. Rev. Nucl. Part. Sci. 44 (1994) 65

\bibitem{Gib75} W.M. Gibson, Ann. Rev. Nucl. Sci. 25 (1975) 465

\bibitem{Gon02} I. Gontchar, M. Morjean and S. Basnary, Europhys. Lett. 57 (2002) 355

\bibitem{Ben02} J. Benlliure, et al., Nucl. Phys. A 700 (2002) 469

\bibitem{Jur04b} B. Jurado et al., Phys. Rev. Lett. 93 (2004) 072501

\bibitem{Jur05b} B. Jurado, et al., Nucl. Phys. A 747 (2005) 14

\bibitem{Jon97} M. de Jong, A. V. Ignatyuk and K.-H. Schmidt, Nucl. Phys. A 613 (1997) 435

\bibitem{Sch97} S. Schmid et al.,  Z. Phys. A. 359 (1997) 27 

\bibitem{Tis05} V. Tishchenko et al., Phys. Rev. Lett. 95 (2005) 162701

\bibitem{Gei92} H. Geissel, et al., Nucl. Instr. and Methods B 70 (1992) 286

\bibitem{Vos95} B. Voss, et al., Nucl. Instr. and Methods A 364 (1995) 150

\bibitem{Pfu94} M. Pf\"utzner, et al. , Nucl. Instr. and Methods B 86 (1994) 213

\bibitem{Sch87} K.-H. Schmidt, et al.,  Nucl. Instr. and Methods A 260 (1987) 287

\bibitem{Ric04} M.V. Ricciardi et al., Phys. Rev. C, accepted

\bibitem{Jun96} A.R. Junghans et al., Nucl. Instr. and Methods A 370 (1996) 312

\bibitem{Tai03} J. Taieb et al., Nucl. Phys. A 724 (2003) 413

\bibitem{Ber03} M. Bernas et al., Nucl. Phys. A 725 (2003) 213

\bibitem{Duf82} J. P. Dufour et al., Nucl. Phys. A 387 (1982) 157c

\bibitem{Enq02} T. Enqvist et al., Nucl. Phys. A 703 (2002) 435

\bibitem{Yar81} Y. Yariv and Z. Fraenkel, Phys. Rev. C 24 (1981) 488

\bibitem{Jun98} A.R. Junghans et al., Nucl. Phys. A 629 (1998) 635

\bibitem{Sie86} A.J. Sierk, Phys. Rev. C 33 (1986) 2039

\bibitem{Ign75} A.V. Ignatyuk, M.G. Itkis, V.N. Okolovich, G.N. Smirekin and A.S. Tishin, Yad. Fiz. 21 (1975) 1185; Sov. J. Nucl. Phys. 21 (1975) 612

\bibitem{Ben98} J. Benlliure et al., Nucl. Phys. A 628 (1998) 458

\bibitem{Kra99} H. J. Krappe, Phys. Rev. C 59 (1999) 2640

\bibitem{Kar03} A. V. Karpov, P. N. Nadtochy, E. G. Ryabov and G. D. Adeev, J. Phys. G 29 (2003) 2365

\bibitem{Sch02} K.-H. Schmidt et al., Nucl. Phys. A 710 (2002) 157

\bibitem{Bou02} A. Boudard, J. Cugnon, S. Leray and C. Volant, Phys. Rev. C 66 (2002) 044615

\bibitem{But91} R. Butsch, D.J. Hofman, C.P. Montoya, P. Paul and M. Thoennessen, Phys. Rev. C 44 (1991) 1515

\bibitem{Ras91} E.M. Rastopchin et al., Yad. Fiz. 53 (1991) 1200 [Sov. J. Nucl. Phys. 53 (1991) 741]

\end{thebibliography}

\end{document}